\documentclass[aps,prb,reprint]{revtex4-1}
\usepackage{amsmath}
\usepackage{amsfonts}
\usepackage{amssymb}
\usepackage{graphics}
\usepackage{float}
\usepackage{graphicx}
\usepackage{epstopdf}
\usepackage[compact]{titlesec}
\usepackage{natbib}
\usepackage{threeparttable}
\usepackage[titletoc,title]{appendix}
\usepackage{lipsum}
\usepackage{varwidth}
\usepackage{tabularx}
\usepackage{color}
\newcommand{\ie}{\textit{i}.\textit{e}., }

\begin{document}
\title{Anomalous photo-thermal effects in multi-layered semi-Dirac black phosphorus}
\author{Parijat Sengupta$^{1}$ and Shaloo Rakheja$^{2}$}
\affiliation{$^{1}$Department of Electrical and Computer Engineering \\ 
Purdue University, West Lafayette, IN 47907, USA.\\
$^{2}$Holonyak Micro and Nanotechnology Laboratory \\
University of Illinois at Urbana-Champaign, Urbana, IL 61801 USA.}

\begin{abstract}
The photo-excitation of multi-layered potassium doped black phosphorus that exists as a gapped semi-Dirac two-dimensional normal insulator (NI), transforms it into a time-reversal symmetry broken Chern insulator (CI). This transition from an NI to CI happens under a high-frequency radiation corresponding to the off-resonant regime of the Floquet theory of periodic perturbations, which here is the impinging light beam. This photo-induced topological CI phase manifests as a finite Berry curvature and forms the basis of anomalous transverse heat flow in presence of a longitudinal temperature gradient. The anomalous variants of the Ettinghausen (EE) and Righi-Leduc effects (RLE), via their respective coefficients are quantitatively analyzed in this work. The strength of anomalous EE and RLE coefficients is a direct outcome of the sum of Berry curvatures over the occupied bands and is shown to drop as the Fermi level is positioned high in the conduction states or deep in the valence region. The drop is attributed to the flipped sign of the Berry curvature in the conduction and valence bands. Additionally, when the Fermi level lies in the band gap, the contribution of the conduction and valence bands may numerically outweigh one another for a much larger value of the coefficients. Finally, we point out how beyond the role of Berry curvature, several laboratory accessible methods can be utilized to modulate the EE and RLE coefficients, including an application of strain, variations in the dopant concentration, and frequency of incident radiation.
\end{abstract}
\maketitle

\vspace{0.3cm}
\section{Introduction}
\vspace{0.3cm}
The Landau theory of phase transitions pivot around the idea of symmetry breaking and explains a diverse set of states of matter including superconductivity and ferromagnetism. Each such state of matter is characterized by a local order parameter~\cite{sethna2006statistical} mirroring a broken symmetry; for instance, the violation of rotational symmetry in the alignment of spins in a ferromagnet. The emergence of phenomena such as the integer and fractional quantum Hall effects, however, fall outside the purview of Landau's symmetry-breaking paradigm and are instead explained by a non-local order parameter. Such states of matter are said to possess topological order~\cite{ando2013topological,qi2011topological} and distinguished by the existence of an invariant and are impervious to external local perturbations. The topological order usually manifests as an observable, chiefly through transport and spectroscopic measurements in topological insulators; for instance, in two-dimensional TIs, it is verified by detecting the topologically protected edge states through their hallmark transport signature~\cite{konig2008quantum,lee2013impurity} while ARPES-like spectroscopy methods~\cite{shan2010effective,hajlaoui2013time} in three-dimensional TIs establish the non-trivial character of the bulk and zero-energy states localized at the boundary. In addition to transport and spectroscopic signatures, it is possible to utilize thermodynamic markers~\cite{quelle2016thermodynamic} to identify topological phase transitions that are evident through an associated set of thermal effects illustrating the underlying broken status of one or both the fundamental symmetries of time reversal and parity; as a case in point, thermodynamic contributions have been shown to be enhanced~\cite{xu2014enhanced,ghaemi2010plane} when edge modes are explicitly considered in thermal Hall measurements. 

In this work, we pursue a closely related paradigm that identifies the emergence of new thermal effects in the event of a topological phase transition. We show how in semi-Dirac black phosphorus (BP) doped with potassium (K) and in a normal insulator (NI) phase (\ie a finite non-inverted band gap), a dynamically engineered topological phase transition is achievable when it is irradiated with a circularly polarized light beam of pre-determined frequency. The layered BP sample irradiated with a high-frequency beam that satisfies the \textit{off}-resonant criterion of the Floquet regime undergoes a band rearrangement and transitions to a Chern insulator (CI) from an NI phase. The onset of such a transition in layered and doped BP under a periodic driving scheme can be revealed via a set of anomalous photo-thermoelectric effects. As a prelude to the proof-of-concept calculations that follow, we make an important remark here about thermodynamic variables such as the total energy and entropy: In general, thermodynamic variables describe the complete system and do not mirror each individual local sub-system that may exist, thus bestowing on them an attribute of non-locality. A thermodynamic variable, therefore, in principle, may serve as a convenient probe to map another intrinsically non-local description - the topological order and its related invariant. Before we proceed further, a brief digression is in order here to describe the choice of material in this paper: Potassium-doped layered BP is characterized by a Dirac semi-metal state~\cite{kim2015observation,baik2015emergence} with a linear and quadratic dispersion along the armchair and zigzag directions, respectively. Potassium as a dopant creates a sufficiently strong internalized electric field to control the band gap that may reduce to zero for a critical doping concentration. The two-dimensional BP is also a well-known thermoelectric~\cite{ling2015renaissance} with a high figure-of-merit (\textit{ZT}); this, in conjunction with the semi-Dirac dispersion presents a rich material platform to examine topologically-driven heat flow. For brevity, note that the target material, \textit{K}-doped multi-layered BP, is simply referred to as BP in rest of the manuscript.

\begin{figure}[t!]
\includegraphics[scale=0.6]{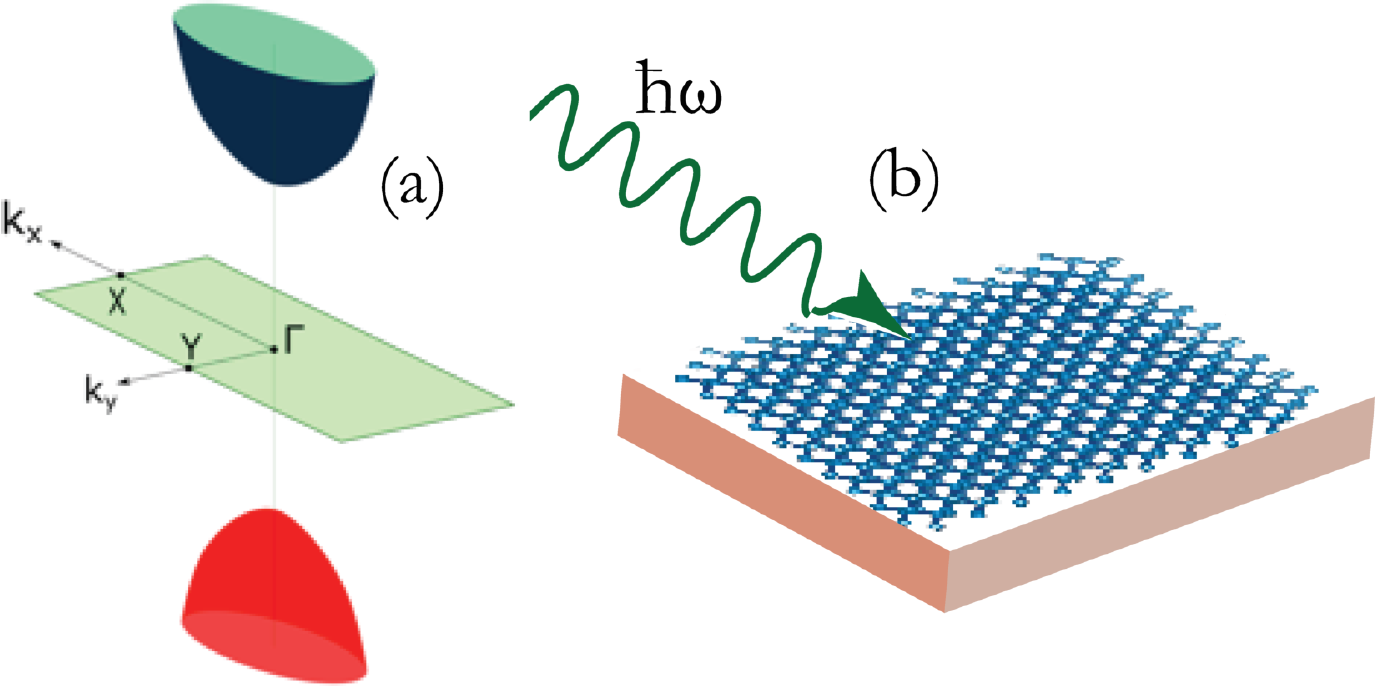}
%\vspace{-0.3cm}
\caption{The schematic on the left panel (a) shows BP with a semi-Dirac dispersion. The dispersion is quadratic and linear along the zigzag and the armchair directions, respectively. The square denotes the surface unit cell with the high-symmetry points marked. The right panel (b) sketches the arrangement wherein an intense light beam of frequency $ \omega $ shines on the BP surface. The interaction between the incident beam and BP alters the intrinsic dispersion and introduces a time reversal symmetry breaking term, which is effectively a transition from a normal insulator (NI) to a Chern insulator (CI). This non-trivial topological phase transformation from NI to CI through a finite Berry curvature drives the anomalous transverse heat flow analyzed in this work.}
\label{sch1}
\vspace{-0.3cm}
\end{figure} 

We begin by defining the characteristic Hamiltonian of the layered BP system and its corresponding set of eigen values and states. The Hamiltonian, using the Floquet theorem, is then transformed under a periodic perturbation (illumination of the BP sample) that gives a time-reversal broken rearranged energy spectrum and a non-trivial band topology. A finite and dynamical Berry curvature serves as a signpost for this topological non-triviality, and forms the genesis for the pair of anomalous transverse heat flow phenomena analyzed in this work: The Ettinghausen (EE) and Righi-Leduc (RLE) effects.~\cite{abrikosov2017fundamentals} Primarily, we observe through generalized expressions for $\alpha_{xy}$, the EE coefficient, and $\kappa_{xy}$, which defines RLE, a quantitative dependence on the Berry curvature $\left(\Omega\left(k\right)\right)$ of the occupied states. A region of momentum space with a large $ \Omega\left(k\right) $ manifests in higher numbers for $ \alpha_{xy} $ and $ \kappa_{xy} $; furthermore, the oppositely directed $\Omega\left(k\right)$ for the conduction and valence states steers these coefficients to exhibit small values when the Fermi level is placed much above or below the band gap region. Additionally, $ \alpha_{xy} $ and $ \kappa_{xy} $ rise with an increase in temperature. We also show that the coefficients can be adjusted through rearrangement of the BP dispersion, utilizing schemes that can be accessed in a laboratory setting via tuning of material properties through application of external strain, modulation doping, and frequency of the photo-excitation.

\vspace{0.3cm}
\section{Electronic structure of layered BP}
\vspace{0.3cm}
We first note the Hamiltonian for a four-layered \textit{K}-doped BP sheet in NI phase as a starting point for further calculations. The effective low-energy Hamiltonian in this case is given as~\cite{baik2015emergence}:
\begin{equation}
\mathcal{H}_{p} = \left(\Delta + \alpha k_{x}^{2}\right)\sigma_{x} + \beta k_{y}\sigma_{y}.
\label{ham1}
\end{equation}
The band gap is $ 2\Delta $. The other terms are defined as $ \alpha = \hbar^{2}/2m^{*} $ and $ \beta = \hbar v_{f} $, where $ m^{*} = 1.42\,m_{e} $ is the effective mass for the parabolic branch and $ v_{f} = 5.6 \times 10^{5} m/s $ is the Fermi velocity along the linear band. As the \textit{K}-doping increases, $ \Delta $ begins to shrink, eventually diminishing to zero. The Hamiltonian, $ \mathcal{H}_{d} $, is therefore just $ \alpha k_{x}^{2}\sigma_{x} + \beta k_{y}\sigma_{y} $. The subscripts \textit{p} and \textit{d} denote pristine and semi-Dirac \textit{K}-doped BP. Diagonalizing Eq.~\ref{ham1}, the eigen energy expressions in the Dirac semi-metal phase is
\begin{equation}
E\left(k\right) = \pm\,\sqrt{\left(\Delta + \alpha k_{x}^{2}\right)^{2} + \left(\beta k_{y}\right)^{2}}.
\label{bpdp}
\end{equation}
The + (-) sign in the energy expressions denote the conduction (valence) state. The dispersion of the doped sample by letting $ \Delta \rightarrow 0 $ in Eq.~\ref{bpdp} clearly points to massless Dirac Fermions along the armchair direction (\textit{y}-axis) while the zigzag axis (\textit{x}-axis) hosts its massive counterpart. The pristine form of this BP dispersion in NI phase is alterable under the influence of a periodic driving insofar as newer topological transitions can be achieved with distinct broken symmetries. In the following section, we outline how irradiating BP with a suitable frequency fundamentally recasts the original dispersion and transforms the NI phase material into a Chern insulator. A Chern insulator is a topological state of matter with integer Hall conductivity in absence of an external magnetic field and the attendant Landau levels. There is no naturally occurring Chern insulator discovered yet, but has been engineered on several two-dimensional lattices.~\cite{minarelli2019engineering,sticlet2012geometrical} 

\vspace{0.3cm}
\section{Periodically-driven semi-Dirac BP Hamiltonian}
\vspace{0.3cm}
\noindent Quantum systems with a periodic Hamiltonian of the form $ \hat{H}\left(t\right) = \hat{H}\left(t + T\right) $ are studied within the framework of the Floquet formalism.~\cite{kok2010introduction,bukov2015universal,kohler2005driven} The time-dependent Hamiltonian is $ \hat{H}\left(t\right) = \hat{H}_{0} + \hat{V}\left(t\right)$, where $ \hat{H}_{0} $ is the stationary part and $ V\left(t\right)= V\left(t + T\right) $ is the time-periodic perturbation. An ansatz of the form $ \Psi_{\alpha}\left(t\right) = \exp\left(-iE_{\alpha}t/\hbar\right)\Phi_{\alpha}\left(t\right) $, where $ \Phi\left(t\right) = \Phi\left(t + T\right) $ (periodic in time) when substituted  in the time-dependent Hamiltonian $ \hat{H}_{F}\left(t\right)\Psi\left(t\right) =  \left(\hat{H}\left(t\right) - i\hbar\partial_{t} \right)\Psi\left(t\right) = 0 $ gives an eigen equation of the form~\cite{tannor2007introduction}
\begin{equation}
\hat{H}_{F}\left(t\right)\Phi_{\alpha}\left(t\right) = E_{\alpha}\Phi_{\alpha}\left(t\right). 
\label{floeq}
\end{equation}
The Floquet theory also tells us that higher modes of the form $ \Phi_{\alpha^{'}}\left(t\right) = \Phi_{\alpha}\left(t\right)\exp\left(in\omega t\right) $ with $ n = 0,\pm 1, \pm 2, ...$ are valid solutions of Eq.~\ref{floeq} but with quasi-energies $ E_{\alpha^{'}} = E_{\alpha} + n\hbar\omega $ and $ T = 2\pi/\omega $. The overall Floquet Hamiltonian $ \left(\hat{H}_{F}\left(t\right)\right) $ which contains the complete class of solutions indexed by the integer $ n $ takes a matrix form
\begin{equation}
\begin{pmatrix}
\ddots & \\
\dots & V_{-2,-3} & H_{0} - 2\omega & V_{-2,-1} & V_{-2,0} & V_{-2,1}  & \dots \\
\dots & V_{-1,-3} & V_{-1,-2} & H_{0} - \omega & V_{-1,0}  & V_{-1,1}  & \dots \\
\dots & V_{0,-3} & V_{0,-2} & V_{0,-1} & H_{0} & V_{0,1}   & \dots \\
\dots & V_{1,-3} & V_{1,-2} & V_{1,-1} & V_{1,0} & H_{0} + \omega  & \dots \\
\quad & \quad & \quad & \quad & \quad & \quad & \quad & \ddots
\end{pmatrix}.
\label{flhamstr}
\end{equation}
In Eq.~\ref{flhamstr}, the \textit{off}-diagonal interaction elements of the matrix are the Fourier transform:~\cite{de2019floquet,eckardt2015high} 
\begin{equation}
V_{mn} = \dfrac{1}{T}\int_{0}^{T}V\left(t\right)\exp\left[-i\left(m-n\right)\omega t\right]dt.
\label{intfour}
\end{equation}

In a two-level system such as the one described by the Hamiltonian in Eq.~\ref{ham1}, the periodic perturbation induced interaction term $\left(V\left(t\right)\right) $, say from a laser source, is of the form $ \sigma \cdot A\left(t\right) $. Here, the time-dependent vector potential is $ A\left(t\right) $. For a single mode (monochromatic) laser source, the vector potential is $ A\left(t\right) = A\left(\cos\omega t, \sin\omega t\right) $. Inserting this sinusoidal form of $ V\left(t\right) $ in Eq.~\ref{intfour}, it is straightforward to notice that the integral in Eq.~\ref{intfour} vanishes unless $ m - n = 0, \pm 1 $. The generalized Floquet Hamiltonian (Eq.~\ref{flhamstr}) then simplifies to a block tri-diagonal form; for example, terms such as $ V_{-2,-3} $, $ V_{-1,-3} $ etc. cease to exist. The finite entries, in addition to the diagonal elements (also called Floquet side-bands)are terms such as $ V_{-1,-2} $. As further simplification, for cases where the light frequency $ \left(\omega\right) $ is much higher than the energy scales of the stationary Hamiltonian, $ H_{p} $, in Eqn.~\ref{ham1}, we can truncate the Floquet matrix retaining just the $ n = 0, \pm 1 $ terms to write~\cite{lopez2015photoinduced}: 
\begin{equation}
\hat{H}_{F, off-resonant} = \hat{H}_{0} + \dfrac{\left[V_{-1}, V_{1}\right]}{\hbar\omega}.
\label{fleq1}
\end{equation}
The terms contained in the anti-commutator in Eqn.~\ref{fleq1} are $ V_{m-n = -1} $ and $ V_{m-n = 1} $ - the Fourier components indicated in Eq.~\ref{intfour}. This truncated Floquet Hamiltonian is commonly referred to as the \textit{off}-resonant approximation.

We use the formalism of \textit{off}-resonant approximation on semi-Dirac BP to derive the attendant changes to the dispersion; to do so, it necessary to first obtain the Fourier components starting from the defining Hamiltonian in Eq.~\ref{ham1}. Assuming an intense periodic driving force which can simply be a laser source irradiates semi-Dirac BP, the Hamiltonian in Eq.~\ref{ham1} transforms to a time-dependent form by applying the Peierls substitution $ \hbar\mathbf{k} \rightarrow \hbar\mathbf{k} - e\mathbf{A}\left(t\right) $. This represents the electromagnetic field-matter coupling via the vector potential $ \mathbf{A}\left(t\right) $. We rewrite Eq.~\ref{ham1} to reflect this time-dependence as~\cite{sengupta2017anisotropy} 
\begin{align}
\hat{H}\left(t\right) &=  \left[\Delta + \dfrac{\alpha}{\hbar}\left(k_{x} + eA_{x}\left(t\right)\right)^{2}\right]\sigma_{x} \notag \\
&+ \dfrac{\beta}{\hbar}\left(k_{y} + eA_{y}\left(t\right)\right)\sigma_{y}.
\label{fleq3}
\end{align}
The time-dependent part is therefore, 
\begin{equation}
\begin{aligned}
V\left(t\right) &= e\alpha/\hbar\left(2k_{x}A_{x}\left(t\right) + eA_{x}^{2}\left(t\right)\right)\sigma_{x} \\
& + \left(e\beta/\hbar\right) A_{y}\left(t\right)\sigma_{y}.
\label{vtd}
\end{aligned}
\end{equation}
For circularly polarized light propagating along $ \hat{z} $, the two vector components are $ \mathbf{A} = A_{0}\left(-\sin\omega t\hat{x}, \cos\omega t\hat{y}\right) $. Here $ A_{0} = E_{0}/\omega $ and the corresponding electric field (via the relation $ \mathbf{E} = -\partial_{t}\mathbf{A} $) is $ \mathbf{E} = E_{0}\left(\cos\omega t\hat{x}, \sin\omega t\hat{y}\right) $. From the time-dependent part, the two Fourier components in Eq.~\ref{fleq1} are $ \left(H_{\eta}, \eta = \pm 1 \right) $ can be evaluated. We have
\begin{equation}
\begin{aligned}
H_{\eta} &= \dfrac{1}{T}\int_{0}^{T}\biggl[\dfrac{\alpha\sigma_{x}}{\hbar}\left(-2A_{0}ek_{x}\sin\omega t + A_{0}^{2}e^{2}\sin^{2}\omega t\right) \\
& + \dfrac{e\beta A_{0}\sigma_{y}}{\hbar}\cos\omega t\biggr]exp\left(i\eta\omega t\right)dt, \\
&=  \dfrac{1}{T}\left[\dfrac{2i\eta \pi e\alpha A_{0}k_{x}\sigma_{x}}{\hbar\omega} + \dfrac{e\pi\beta A_{0}\sigma_{y}}{\hbar\omega}\right].
\label{comms}
\end{aligned}
\end{equation}
The commutator in Eq.~\ref{fleq1} using the Pauli relation, $ \left[\sigma_{x},\sigma_{y}\right] = 2i\sigma_{z} $, and the result from Eq.~\ref{comms} is
\begin{equation}
\dfrac{\left[H_{-1}, H_{1}\right]}{\hbar\omega} = -\dfrac{2}{\hbar\omega}\left(eA_{0}v_{f}\right)^{2}\left(\alpha/\beta\right)k_{x}\sigma_{z}.
\label{commsf}
\end{equation}
The rearranged Hamiltonian (superscript $ F $ indicates Floquet-modified) for four-layered BP using Eq.~\ref{commsf} is
\begin{equation}
H_{p}^{F} = \left(\Delta + \alpha k_{x}^{2}\right)\sigma_{x} + \beta k_{y}\sigma_{y}\,+ \mathcal{F}k_{x}\sigma_{z}.
\label{flham}
\end{equation}
Here, $ \mathcal{F} = -2\left(\hbar\omega\right)^{-1}\left(eA_{0}v_{f}\right)^{2}\left(\alpha/\beta\right) $. The dispersion by diagonalizing Eq.~\ref{flham} is
\begin{equation}
E^{F}\left(k\right) =  \pm\,\sqrt{\mathcal{F}^{2}k_{x}^{2} + \left(\Delta + \alpha k_{x}^{2}\right)^{2} + \left(\beta k_{y}\right)^{2}}.
\label{fldp}
\end{equation}
The term $ eA_{0}v_{f} $ has unit of energy.

\begin{figure}[t!]
\includegraphics[scale=0.55]{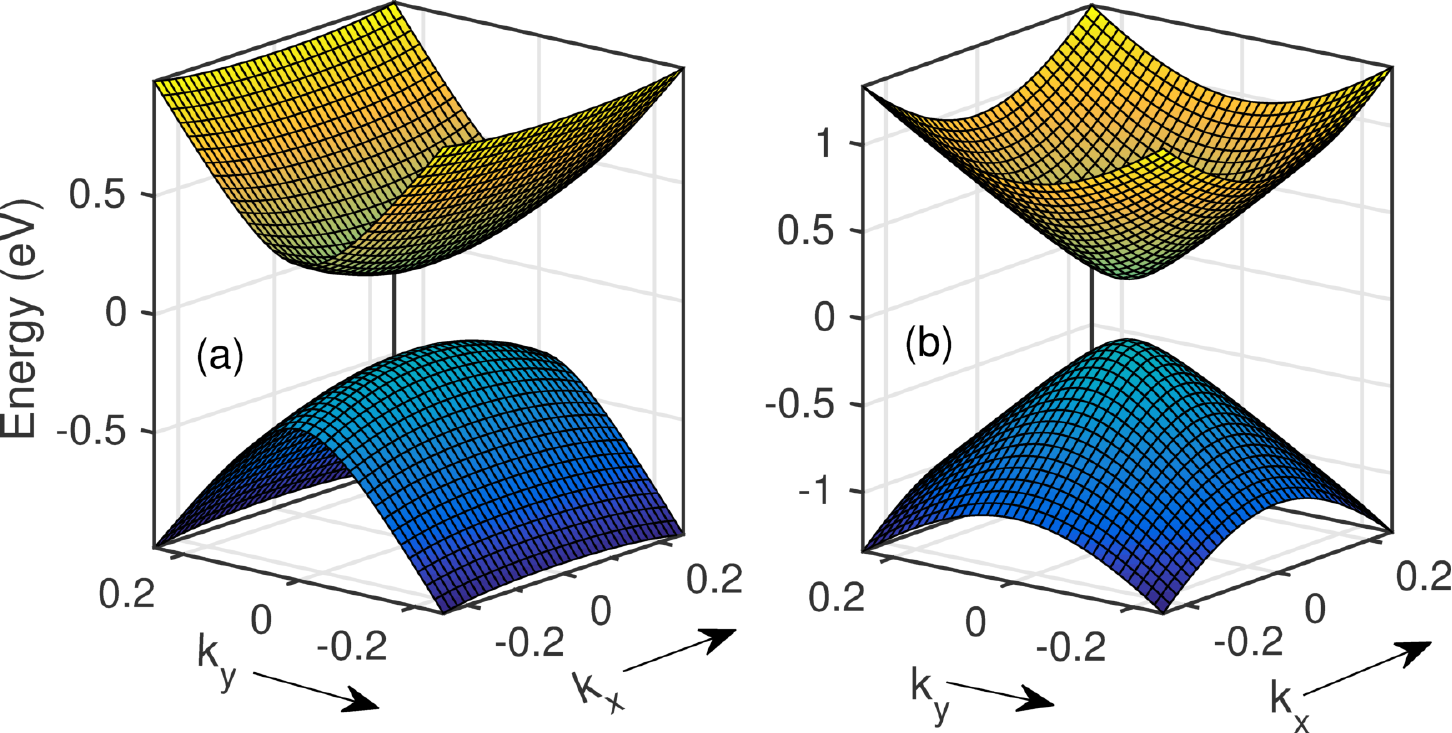}
\caption{TThe numerically calculated electronic dispersion of semi-Dirac BP in shown in the left panel (a) where the bands appear distinctly parabolic. However, under an intense light beam the bands are rearranged and acquires a linear character (b) as revealed in the funnel-like shape. For visual clarity, we have artificially set the $ eA_{0}v_{f} $ term in the Floquet contribution (Eq.~\ref{commsf}) of the beam to a large value of $ 5.0\, eV $ while the frequency corresponds to an optical source of energy $ 7.0\, eV $. The $k$-components are in $\AA^{-1}$.}
\label{disp}
\vspace{-0.3cm}
\end{figure} 

\vspace{0.3cm}
\subsection{Chern insulator phase}
\vspace{0.3cm}
The assertion that semi-Dirac \textit{K}-doped BP transitions to a Chern insulator from a normal insulator can be seen by a simple examination of the additional introduced term $ \mathcal{F}k_{x}\sigma_{z} $ in the \textit{off}-resonant approximation. Setting aside the $ \omega $-dependent $ \mathcal{F} $ which is a number, we consider the product $ k_{x}\sigma_{z} $ for the underlying symmetries. Under the action of the parity operator $ \mathcal{P} $, we have $ \mathbf{k} \overset{\mathcal{P}}{\rightarrow} -\mathbf{k} $ and $ \sigma_{z} \overset{\mathcal{P}}{\rightarrow} -\sigma_{z} $; therefore, this term preserves parity. Note that this $ \mathcal{P} $ operation takes sub-lattice 1 into 2. Similarly, under the time-reversal symmetry operator $ \mathcal{T} $, the momentum vector changes from $ \mathbf{k} \overset{\mathcal{T}}{\rightarrow} -\mathbf{k} $ while  $ \sigma_{z} \overset{\mathcal{T}}{\rightarrow} \sigma_{z} $. The sub-lattice degree of freedom $ \sigma_{z} $ remains unaffected when acted upon by the $ \mathcal{T} $ operator. We can therefore write $ k_{x}\sigma_{z} \overset{\mathcal{T}}{\rightarrow} -k_{x}\sigma_{z} $, from which we clearly infer a loss of time reversal invariance. A system that loses time reversal symmetry without an external magnetic field is a manifestation of the Haldane model, the archetype Chern insulator.~\cite{haldane1988model}. For our case, this is an instance of a Floquet Chern insulator (FCI) dynamically prepared under the influence of a periodic perturbation. Note that the semi-Dirac BP is an FCI when the \textit{off-resonant} approximation remains a valid simplification of the generalized Floquet Hamiltonian. For completion, we may add here that dynamical generation of topological Hamiltonian have been reported before, for example, engineered FCIs.~\cite{d2015dynamical} It is also pertinent to emphasize that layered BP possesses no intrinsic topological order, but by virtue of generation of a photo-aided dynamical topological Hamiltonian, a quantum phase transition to FCI occurs.

A time reversal symmetry broken system must have a finite Berry curvature $\left(\Omega\left(k\right)\right) $. In our case, a finite $ \Omega\left(k\right) $ is proven using the prescribed formalism for a two-level system. For a quantum well, which is a two-dimensional system whose Hamiltonian (disregarding the quadratic component that does not- contribute to $ \Omega\left(k\right)$ ) is expressible as $ H\left(k\right) = \mathbf{d\left(k\right)}\cdot \mathbf{\sigma} $, the Berry curvature is defined as~\cite{fruchart2013introduction} 
\begin{equation}
\Omega_{\mu\nu} = \dfrac{1}{2}\varepsilon_{\alpha\beta\gamma}\hat{d}_{\alpha}\left(\mathbf{k}\right)\partial_{k_{\mu}}\hat{d}_{\beta}\left(\mathbf{k}\right)\partial_{k_{\nu}}\hat{d}_{\gamma}\left(\mathbf{k}\right),
\label{berry}
\end{equation}
where $ \hat{\mathbf{d}}\left(\mathbf{k}\right) = \dfrac{\mathbf{d\left(\mathbf{k}\right)}}{d\left(k\right)} $. Applying this formalism in the case of the semi-Dirac BP Hamiltonian (Eq.~\ref{ham1}) and noting that $ d_{x} = \alpha k_{x}^{2} $, $ d_{y} = \beta k_{y} $, and $ d_{z} = \mathcal{F}k_{x} $, we can write down the following expression for $ \Omega\left(k\right) $
\begin{equation}
\Omega\left(k\right) = \mp\dfrac{\alpha\beta\mathcal{F}k_{x}^{2}}{\left(\mathcal{F}^{2}k_{x}^{2} + \left(\Delta + \alpha k_{x}^{2}\right)^{2} + \left(\beta k_{y}\right)^{2}\right)^{3/2}}.
\label{bc}
\end{equation}
The upper (lower) sign is for the conduction (valence) bands. We used symmetry arguments above to point out how the Floquet-induced term, when semi-Dirac BP is irradiated, breaks time-reversal symmetry (TRS). The loss of TRS is also explicitly seen from the $\Omega\left(k\right)$ expression in Eq.~\ref{bc}. This can be easily gauged from the condition that for systems with TRS, $\Omega\left(k\right)$ is an odd function of the momentum vector. In our case, it is however easy to see $\Omega\left(k\right) = \Omega\left(-k\right)$, and therefore TRS does not exist.
\begin{figure}[t!]
\includegraphics[scale=0.7]{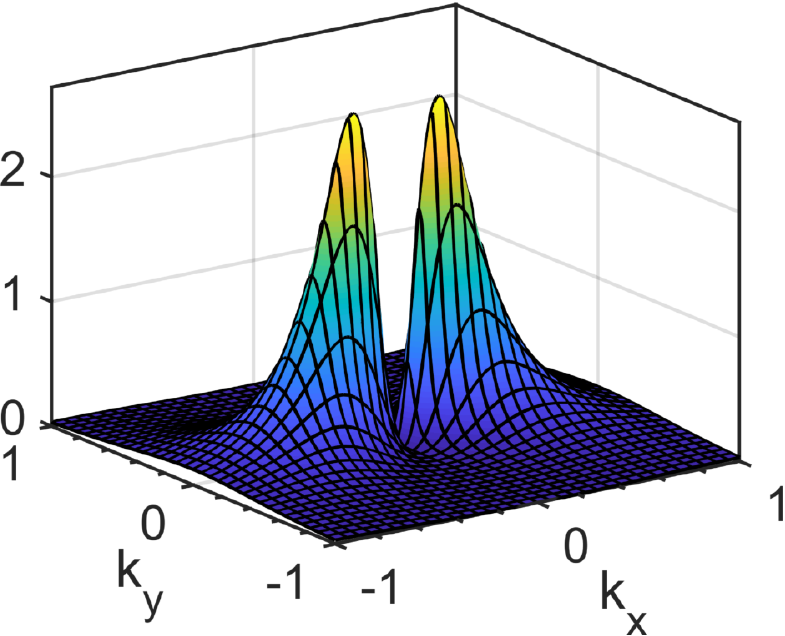}
%\vspace{-0.3cm}
\caption{The Berry curvature $\left(\Omega\left(k\right)\right)$ as shown in Eq.~\ref{bc} for a sample of BP under a photo-excitation is plotted in this figure. As is evident from the expression for $ \Omega\left(k\right)$, it is positive everywhere and peaks in regions of momentum space where the energy is the least, which happens to be around the top of the valence and bottom of the conduction bands. The $ \Omega\left(k\right)$ is also an even function of the momentum vector $\left(\Omega\left(k\right) = \Omega\left(-k\right)\right)$ and therefore indicates the absence of time-reversal symmetry.}
\vspace{-0.5cm}
\label{ber}
\end{figure}

This non-trivial Berry curvature imparts an anomalous velocity to Bloch electrons skewing their trajectory, the effects of which are noticeable transverse-Hall like electric and thermal currents when external magnetic fields are absent. In the following section, we examine a set of thermoelectric effects that are well-understood from a macroscopic standpoint and describe the interplay between electric and thermal currents in presence of an external magnetic field $\left(\mathbf{B}\right)$. The Berry curvature is the substitutive microscopic counterpart of an external $ \mathbf{B} $ field leading to the production of finite thermal and voltage gradients.

\vspace{0.3cm}
\section{Anomalous thermal effects}
\vspace{0.3cm}
A variety of phenomena may occur when current flows in the simultaneous presence of electric and temperature gradients and an external magnetic field. We follow the notation of Landau and Lifshitz~\cite{landau2013electrodynamics} to identify the contribution to the overall heat current arising from each such phenomenon. The heat current is succinctly expressed as
\begin{equation}
J_{q} = \alpha T J_{e} - \kappa\triangledown T + NT\left(\mathbf{H} \times J_{e}\right) + L\mathbf{H} \times \triangledown T.
\label{heq} 
\end{equation}

\begin{figure}[t!]
\includegraphics[scale=0.55]{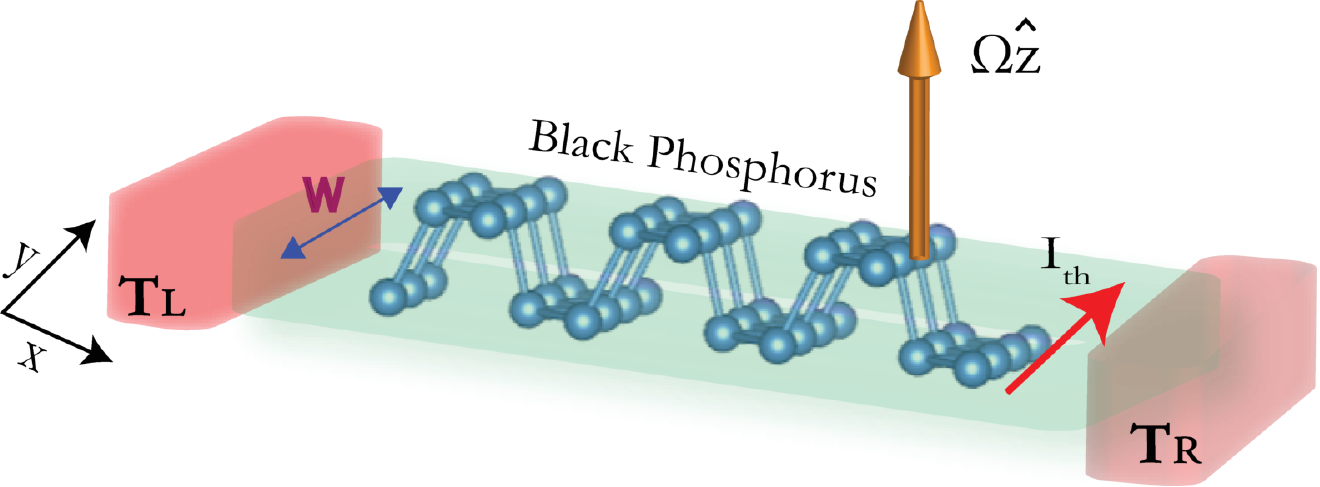}
%\vspace{-0.3cm}
\caption{The thermal Hall effect (RLE) is schematically shown here. The two-dimensional BP is clamped between contacts at ends maintained at different temperatures. In a classical arrangement, the setup as depicted above is placed in a constant magnetic flux density. A temperature gradient (and heat flow marked by the red arrow) develops transverse to the longitudinal difference of $ T_{L} - T_{R}$ and the magnetic field. The magnetic field, in case of an anomalous variant of the RLE, is substituted by the Berry curvature $\left(\Omega\left(k\right)\right)$. The RLE is independent of the size of the sample. Note that when the longitudinal heat flow of RLE is replaced by an electric current, a similar transverse temperature gradient is observed; this is known as the Ettinghausen effect.}
\vspace{-0.3cm}
\label{ansch}
\end{figure}

In Eq.~\ref{heq}, the heat current is $ J_{q} $, the electric current density is $ J_{e} $, while the terms in the order shown represent the Peltier effect, Fourier's law of thermal conduction, the Ettingshausen (EE) and Righi-Leduc (RLE) effects, respectively. In this section, we quantitatively analyze and discuss the anomalous variants of EE and RLE in the context of layered BP that acquires a finite Berry curvature as it transitions to a Floquet Chern insulator. Before we begin, it is appropriate to mention that EE and RLE constitute the transverse flow of thermal currents and both arise from the curvature of thermal electrons under a magnetic field. For our case, the magnetic field is the momentum dependent $ \Omega\left(k\right) $. In presence of a temperature gradient, the anomalous thermal conductivity, also known as the thermal Hall effect or the RLE is formally defined via the relation $ j_{q,y} = -\kappa_{xy}\nabla_{x}T $. The heat current along \textit{y}-axis is $ j_{q,y} $, and transverse to a temperature gradient vector aligned to the \textit{x}-axis (Fig.~\ref{ansch}. This transverse heat flow is dependent on $\Omega\left(k\right)$ and given as~\cite{yokoyama2011transverse,bergman2010theory}
\begin{equation}
\begin{aligned}
\kappa_{xy} &= \dfrac{k_{B}^{2}T}{h}\int\dfrac{\mathbf{d}^{2}k}{4\pi^{2}}\sum_{\tau}\Omega_{\tau}\left(k\right)\biggl[\dfrac{\pi^{2}}{3} + k_{B}T\left(E - \mu\right)f \\
&- ln^{2}\left(1 - f\right) - 2Li_{2}\left(1 - f\right)\biggr].
\label{thber}  
\end{aligned}
\end{equation}
In Eq.~\ref{thber}, $ Li_{n}\left(x\right) = \sum_{m = 1}^{\infty}\dfrac{x^{m}}{m^{n}} $ is the poly-logarithmic function. The symbol $ f $ denotes the equilibrium Fermi distribution. A similar form of transverse heat flow is the Ettinghausen effect (more precisely, a transverse thermal gradient produced by a longitudinal charge current in presence of a perpendicular magnetic field), which is characterized by the off-diagonal Peltier coefficient and given as~\cite{xiao2006berry}
\begin{equation}
\begin{aligned}
\alpha_{xy} &= -\dfrac{k_{B}e}{h}\int\dfrac{\mathbf{d}^{2}k}{4\pi^{2}}\sum_{\tau}\Omega_{\tau}\left(k\right)\biggl[\beta\left(E - \mu\right)f \\
& + ln\left(1 + exp\left(-\beta\left(E - \mu\right)\right)\right)\biggr].
\label{pelber}  
\end{aligned}
\end{equation}

\begin{figure}[t!]
\includegraphics[scale=0.71]{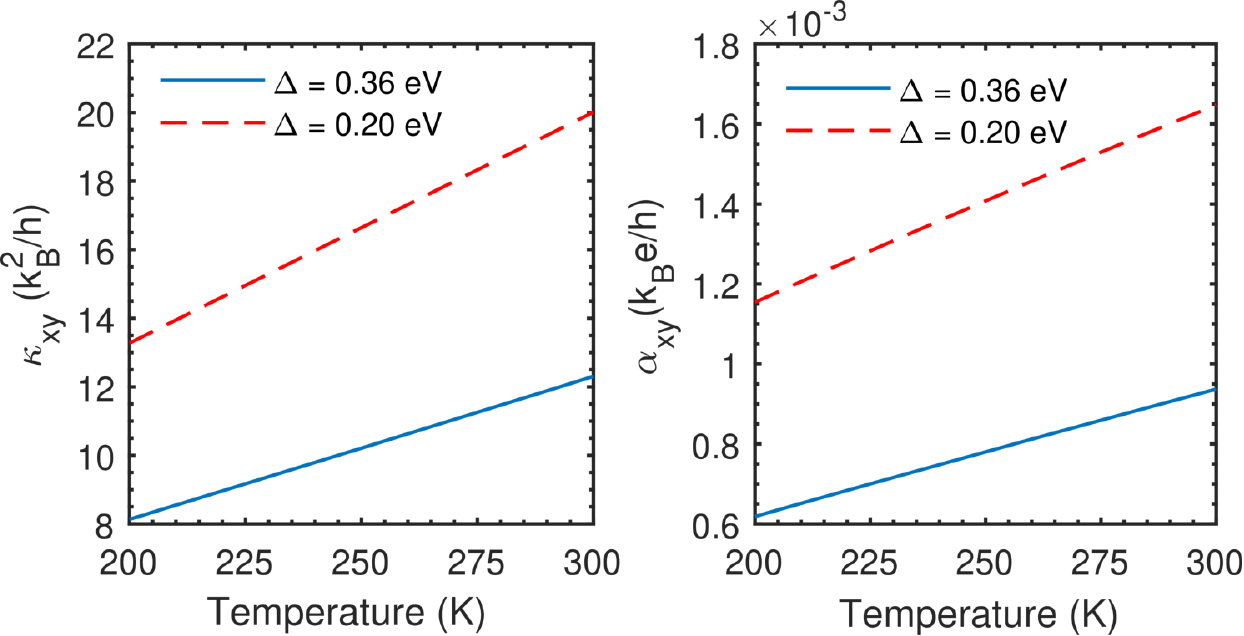}
\centering
%\vspace{-0.5cm}
\caption{The numerically determined coefficients for the anomalous Righi-Leduc $\left(\alpha_{xy}\right)$ and Ettinghausen $\left(\kappa_{xy}\right)$ effects are plotted here for a range of temperatures. The Fermi level for the two band gaps considered in these calculations is placed at the respective bottom of the conduction band $\left(\Delta/2\right)$. Clearly, both coefficients, as discussed in the text, show a rise with temperature as more states are occupied (smearing of the Fermi distribution function) allowing for a larger flow of thermal current. In preparing these plots, we numerically integrated Eqs.~\ref{thber} and ~\ref{pelber} over a circular patch of radius $ 0.25\,\AA^{-1} $ in momentum space. Note that $ \alpha_{xy} $ and $ \kappa_{xy} $ are plotted in units of $ k_{B}^{2}/h $ and $ k_{B}e/h $, respectively.}
\vspace{-0.4cm}
\label{kpal}
\end{figure}

The two transverse heat coefficients $\left(\kappa_{xy},\alpha_{xy}\right)$ are numerically computed for a range of temperatures. The left (right) panel in Fig.~\ref{kpal} shows the RLE (EE) conductivity coefficients. In preparing these plots, the laser power (the term $ eA_{0}v_{f} $ in Eq.~\ref{fldp}) was set to $ 3.0\,eV $ and the beam energy is $ \hbar\omega = 8.0\,eV $. The effective mass and Fermi velocity (parameters that describe the pristine BP Hamiltonian in Eq.~\ref{ham1}) are identical to those listed in the caption of Fig.~\ref{disp}. Additionally, the coefficients have been calculated for two band gaps (recall that in layered and K-doped BP, band gaps are adjustable) and the Fermi level in each case has been aligned to the bottom of the conduction band. For clarity, the bottom of the conduction band is at $ \Delta/2 $ while the top of the valence band sits at $ -\Delta/2 $ for an effective gap of $ \Delta $. For calculations that follow, the RLE (the thermal Hall) and Ettinghausen coefficients, $ \kappa_{xy} $ and $ \alpha_{xy} $, are normalized in units of $ k_{B}^{2}/h $ and $ k_{B}e/h $, respectively. First of all observe from the defining equations for these coefficients (Eqs.~\ref{thber} and ~\ref{pelber}) that the Fermi level and the Berry curvature $\left(\Omega\left(k\right)\right)$ are key contributors that influence any quantitative evaluation. With this information, we immediately notice from the plots that both coefficients increase with temperature (set in the range $ 200\, K \leqslant T \leqslant 300\,K $) and reconcile to this profile of the curves by simply noting that at higher temperatures the thermal broadening smears the Fermi function over a large range of electronic states, and enhances the overall scope of thermal conduction. 

In conjunction with this temperature dependence, the numerical analysis also showed that $\kappa_{xy}$ and $\alpha_{xy}$ exhibit a drop in magnitude as the Fermi level $\left(\mu\right)$ is moved higher in the conduction bands or placed deep in the valence region; we consider two scenarios here: 1) The $\left(\mu\right)$ is placed high in the conduction bands $\left(\mu > \Delta\right)$. In the two-band model considered here, we work with the lowest conduction band and highest valence band. Both bands are filled for this position of $ \mu $ and at a finite temperature contribute to the thermal current. As pointed above, the anomalous variants of the Ettinghausen and Righi-Leduc effects receive contributions from the Berry curvature $\left(\Omega\left(k\right)\right)$; their respective coefficients (Eqs.~\ref{thber} and ~\ref{pelber}) quantitatively depend on the sum of $ \Omega\left(k\right) $ over the CB and VB states. The sign of $\Omega\left(k\right)$, however, if we recall, flips between the conduction (CB) and valence (VB) bands. It is therefore easy to see that while CB and VB contributions to the strength of the coefficients are finite, their combined role on account of the flipped sign of $ \Omega\left(k\right)$ diminishes the overall strength. 

\begin{figure}[t!]
\includegraphics[scale=0.70]{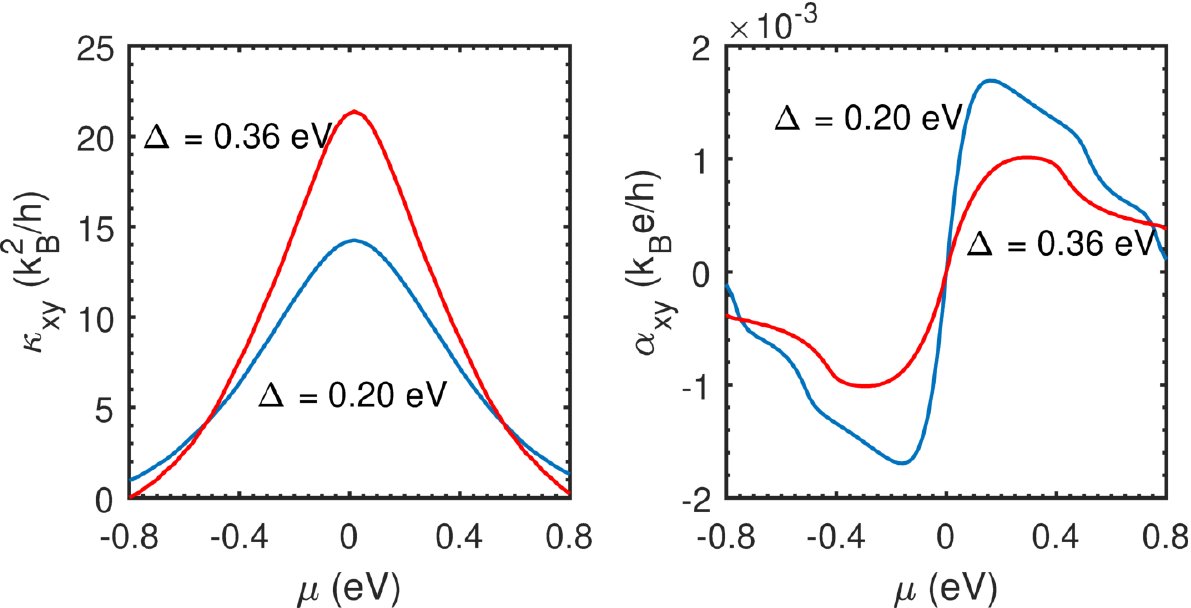}
\centering
%\vspace{-0.3cm}
\caption{The coefficients characterizing the anomalous Righi-Leduc $\left(\alpha_{xy}\right)$ and Ettinghausen $\left(\kappa_{xy}\right)$ are computed for several values of the Fermi level $\left(\mu\right)$. The temperature in these calculations is set to $ T = 300\,K $. The coefficients fall when $\mu $ is pushed up in the conduction band region $\left(\mu > \Delta\right)$ or lies far down in the valence states $\left(-\mu < -\Delta\right)$. In the intermediate region of $ -\Delta/2 < \mu < \Delta/2 $, the coefficients attain higher values. Additionally, notice that $ \alpha_{xy} $ remains positive for all values of $ \mu $ while $ \kappa_{xy} $ changes sign as the Fermi level dives into the valence region. The numerical integration of Eqs.~\ref{thber} and ~\ref{pelber} are identical to that noted in caption of Fig.~\ref{kpal}.}
\vspace{-0.3cm}
\label{kamu}
\end{figure}

The second scenario deals with $ \mu $ lying low in the valence bands $\left(-\mu < -\Delta\right)$. Here, the Fermi function for both CB and VB are close to zero and `empty' states do not support a thermal current flow. As a concrete illustration of this qualitative elucidation, we plot in Fig.~\ref{kamu} the coefficient profiles for a range of $ \mu $ values; as expected, out assertion is indeed borne out as both coefficients droop to the zero mark when $ \mu $ goes deep in the conduction or valence region. The two extreme $ \mu $ points corresponding to the scenarios described are $ -0.8\,eV  \text{and}\,  0.8\,eV $. Additionally, notice that $ \alpha_{xy} $ switches sign as $ \mu $ changes polarity while $ \kappa_{xy} $ remains positive throughout. 

Lastly, for cases where $\mu$ is placed in the band gap, it is possible to ascertain the magnitude of contribution of band specific $ \Omega\left(k\right)$ to the EE and RLE coefficients. We can qualitatively analyze by noticing that the finite temperature induces a certain degree of overlap of the conduction and valence bands with $ \mu $; the band that smears higher (bringing in more states) over $ \mu $ leaves a stronger imprint on the anomalous thermal phenomena via its corresponding $ \Omega\left(k\right) $ values. This essentially implies that either the CB or VB would assume a more dominant role and precludes an almost exact cancellation like situation that happens for $ \mu > \Delta $. Finally, the $ \Omega\left(k\right) $ is concentrated close to the band touching singular point $\left(k = 0\right)$ and the coefficients therefore attain their highest values in its vicinity. 

The tunability of the coefficients have until now focused on the Berry curvature, temperature, and position of Fermi level. Before we close, it is worthwhile to draw attention to the role of dispersion of BP in determining the quantum of transverse thermal current. The plots in Fig.~\ref{kpal} show how for a lower band gap the coefficients rise, which indicate that such alterations to dispersion features may serve as supplementary controls for transverse heat flow. We note here that an illustration of such `control' on the circular dichroism was analytically demonstrated using strained graphene.~\cite{rakheja2016tuning} Furthermore, as is well understood, the band gap, particularly in case of BP can be varied via an external electrostatic gate~\cite{li2017tunable} or the \textit{in situ} electric field offered by appropriate set of dopants. A mechanical strain that gives pseudomorphic lattice deformation and impacts the electron effective mass can also significantly reshape the dispersion embodied through an enlargement or shrinking of the band gap and a new set of eigen states. This clearly rearranges the plots in Fig.~\ref{kpal} via their respective equations (Eqs.~\ref{thber} and ~\ref{pelber}).

\begin{figure}[t!]
\includegraphics[scale=0.68]{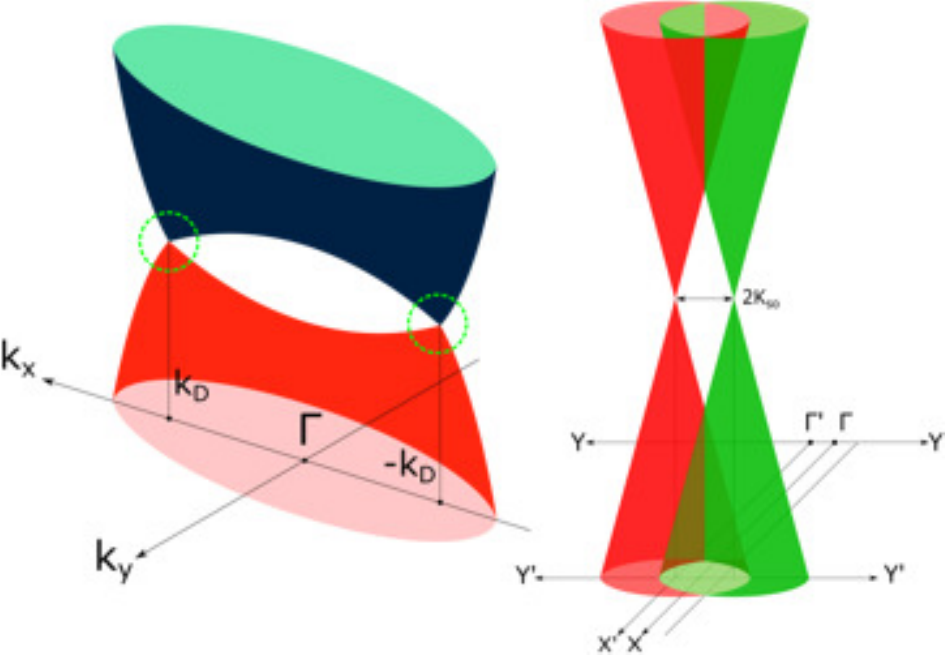}
%\vspace{-0.3cm}
\caption{The gapped BP can turn into an anisotropic Dirac semi-metal as the potassium dopant concentration increases until a threshold value is reached. The left panel shows the band gap closing at two circled nodal points. In the right panel, beyond the dopant threshold concentration, the band gap of the Dirac semi-metal inverts and linear bands cross at two distinct nodes giving rise to a topological insulator (TI). The non-trivial transition to a TI, distinct from the photo-aided conversion of normal to Chern insulator presented in this work, will also carry a finite Berry curvature and be a likely candidate to test the family of anomalous thermal phenomena, including the Ettinghausen and Righi-Leduc effects.}
\vspace{-0.3cm}
\label{dber}
\end{figure}   

\vspace{0.3cm}
\section{Concluding Remarks}
\vspace{0.3cm}
We presented the transition of layered BP from a normal insulator to the Chern insulator phase when irradiated with high-frequency beams that satisfy the \textit{off-resonance} condition in a Floquet analysis of photo-excitation in materials. This topologically delineated transition imparts a non-zero Berry curvature to BP in the Chern insulator phase, which consequently induces anomalous transverse heat flow in presence of a temperature gradient. The coefficients of two such phenomena - the Ettinghausen and Righi-Leduc effect - were examined here. It is useful to note here that while irradiating BP achieves the transition to a Chern insulator phase, similar transitions are also possible by tuning the potassium dopant concentration insofar as the original gapped $\left(\Delta\right)$ semi-Dirac system changes to a topological insulator (TI) with linear dispersion and inverted (negative $\Delta $) bands. The band gap closing and the subsequent inversion is attributable to dopant concentration guided giant Stark effect.~\cite{liu2017gate} The sketch in Fig.~\ref{dber} illustrates this transformation beginning with semi-Dirac BP whose band gap has reduced to zero; a further increase in the $ K $-dopant concentration sets up the TI structure. Note that the anomalous EE and RLE can still be observed in the BP in TI phase, though quantitatively different as the eigen energies and $ \Omega\left(k\right) $ in this case are from a Hamiltonian dissimilar to Eq.~\ref{ham1}.

While anomalous Nernst and Righi-Leduc effects in the context of entropy flow have been studied before~\cite{li2017anomalous} in the noncollinear antiferromagnet Mn$_{3}$Sn, BP, widely recognized a promising thermoelectric material - often noticeable via a large Seebeck coefficient~\cite{flores2015thermoelectric,saito2016gate} - may serve as a suitable platform that combines its in-plane anisotropy and magnetic fields to control the flow of transverse thermal currents through experimentally accessible schemes including the frequency and power of the impinging radiation and mechanical strain to BP. It however remains to be seen (deferred to a future work) if, like Mn$_{3}$Sn, the well-known correlation between the electric and thermal conductivities succinctly expressed by the Wiedemann-Franz law holds for the anomalous coefficients of BP and their ratio hovers around the Lorenz number. In closing, we may also take note of the fact that in absence of external magnetic fields, the anomalous EE and RLE may further augment the figure-of-merit of thermoelectric BP devices and form the basis of thermal modulators whose direction of flow of heat current can be adjusted through simple alterations to the band profile.

%\bibliography{bibfile} 

\begin{thebibliography}{37}
\expandafter\ifx\csname natexlab\endcsname\relax\def\natexlab#1{#1}\fi
\expandafter\ifx\csname bibnamefont\endcsname\relax
  \def\bibnamefont#1{#1}\fi
\expandafter\ifx\csname bibfnamefont\endcsname\relax
  \def\bibfnamefont#1{#1}\fi
\expandafter\ifx\csname citenamefont\endcsname\relax
  \def\citenamefont#1{#1}\fi
\expandafter\ifx\csname url\endcsname\relax
  \def\url#1{\texttt{#1}}\fi
\expandafter\ifx\csname urlprefix\endcsname\relax\def\urlprefix{URL }\fi
\providecommand{\bibinfo}[2]{#2}
\providecommand{\eprint}[2][]{\url{#2}}

\bibitem[{\citenamefont{Sethna et~al.}(2006)}]{sethna2006statistical}
\bibinfo{author}{\bibfnamefont{J.}~\bibnamefont{Sethna}} \bibnamefont{et~al.},
  \emph{\bibinfo{title}{Statistical mechanics: entropy, order parameters, and
  complexity}}, vol.~\bibinfo{volume}{14} (\bibinfo{publisher}{Oxford
  University Press}, \bibinfo{year}{2006}).

\bibitem[{\citenamefont{Ando}(2013)}]{ando2013topological}
\bibinfo{author}{\bibfnamefont{Y.}~\bibnamefont{Ando}},
  \bibinfo{journal}{Journal of the Physical Society of Japan}
  \textbf{\bibinfo{volume}{82}}, \bibinfo{pages}{102001}
  (\bibinfo{year}{2013}).

\bibitem[{\citenamefont{Qi and Zhang}(2011)}]{qi2011topological}
\bibinfo{author}{\bibfnamefont{X.-L.} \bibnamefont{Qi}} \bibnamefont{and}
  \bibinfo{author}{\bibfnamefont{S.-C.} \bibnamefont{Zhang}},
  \bibinfo{journal}{Reviews of Modern Physics} \textbf{\bibinfo{volume}{83}},
  \bibinfo{pages}{1057} (\bibinfo{year}{2011}).

\bibitem[{\citenamefont{K{\"o}nig et~al.}(2008)\citenamefont{K{\"o}nig,
  Buhmann, W.~Molenkamp, Hughes, Liu, Qi, and Zhang}}]{konig2008quantum}
\bibinfo{author}{\bibfnamefont{M.}~\bibnamefont{K{\"o}nig}},
  \bibinfo{author}{\bibfnamefont{H.}~\bibnamefont{Buhmann}},
  \bibinfo{author}{\bibfnamefont{L.}~\bibnamefont{W.~Molenkamp}},
  \bibinfo{author}{\bibfnamefont{T.}~\bibnamefont{Hughes}},
  \bibinfo{author}{\bibfnamefont{C.-X.} \bibnamefont{Liu}},
  \bibinfo{author}{\bibfnamefont{X.-L.} \bibnamefont{Qi}}, \bibnamefont{and}
  \bibinfo{author}{\bibfnamefont{S.-C.} \bibnamefont{Zhang}},
  \bibinfo{journal}{Journal of the Physical Society of Japan}
  \textbf{\bibinfo{volume}{77}}, \bibinfo{pages}{031007}
  (\bibinfo{year}{2008}).

\bibitem[{\citenamefont{Lee et~al.}(2013)\citenamefont{Lee, Liu, Chang, and
  Shen}}]{lee2013impurity}
\bibinfo{author}{\bibfnamefont{H.-H.} \bibnamefont{Lee}},
  \bibinfo{author}{\bibfnamefont{J.-Y.} \bibnamefont{Liu}},
  \bibinfo{author}{\bibfnamefont{C.-R.} \bibnamefont{Chang}}, \bibnamefont{and}
  \bibinfo{author}{\bibfnamefont{S.-Q.} \bibnamefont{Shen}},
  \bibinfo{journal}{Physical Review B} \textbf{\bibinfo{volume}{88}},
  \bibinfo{pages}{195149} (\bibinfo{year}{2013}).

\bibitem[{\citenamefont{Shan et~al.}(2010)\citenamefont{Shan, Lu, and
  Shen}}]{shan2010effective}
\bibinfo{author}{\bibfnamefont{W.-Y.} \bibnamefont{Shan}},
  \bibinfo{author}{\bibfnamefont{H.-Z.} \bibnamefont{Lu}}, \bibnamefont{and}
  \bibinfo{author}{\bibfnamefont{S.-Q.} \bibnamefont{Shen}},
  \bibinfo{journal}{New Journal of Physics} \textbf{\bibinfo{volume}{12}},
  \bibinfo{pages}{043048} (\bibinfo{year}{2010}).

\bibitem[{\citenamefont{Hajlaoui et~al.}(2013)\citenamefont{Hajlaoui,
  Papalazarou, Mauchain, Jiang, Miotkowski, Chen, Taleb-Ibrahimi, Perfetti, and
  Marsi}}]{hajlaoui2013time}
\bibinfo{author}{\bibfnamefont{M.}~\bibnamefont{Hajlaoui}},
  \bibinfo{author}{\bibfnamefont{E.}~\bibnamefont{Papalazarou}},
  \bibinfo{author}{\bibfnamefont{J.}~\bibnamefont{Mauchain}},
  \bibinfo{author}{\bibfnamefont{Z.}~\bibnamefont{Jiang}},
  \bibinfo{author}{\bibfnamefont{I.}~\bibnamefont{Miotkowski}},
  \bibinfo{author}{\bibfnamefont{Y.}~\bibnamefont{Chen}},
  \bibinfo{author}{\bibfnamefont{A.}~\bibnamefont{Taleb-Ibrahimi}},
  \bibinfo{author}{\bibfnamefont{L.}~\bibnamefont{Perfetti}}, \bibnamefont{and}
  \bibinfo{author}{\bibfnamefont{M.}~\bibnamefont{Marsi}},
  \bibinfo{journal}{The European Physical Journal Special Topics}
  \textbf{\bibinfo{volume}{222}}, \bibinfo{pages}{1271} (\bibinfo{year}{2013}).

\bibitem[{\citenamefont{Quelle et~al.}(2016)\citenamefont{Quelle, Cobanera, and
  Smith}}]{quelle2016thermodynamic}
\bibinfo{author}{\bibfnamefont{A.}~\bibnamefont{Quelle}},
  \bibinfo{author}{\bibfnamefont{E.}~\bibnamefont{Cobanera}}, \bibnamefont{and}
  \bibinfo{author}{\bibfnamefont{C.~M.} \bibnamefont{Smith}},
  \bibinfo{journal}{Physical Review B} \textbf{\bibinfo{volume}{94}},
  \bibinfo{pages}{075133} (\bibinfo{year}{2016}).

\bibitem[{\citenamefont{Xu et~al.}(2014)\citenamefont{Xu, Gan, and
  Zhang}}]{xu2014enhanced}
\bibinfo{author}{\bibfnamefont{Y.}~\bibnamefont{Xu}},
  \bibinfo{author}{\bibfnamefont{Z.}~\bibnamefont{Gan}}, \bibnamefont{and}
  \bibinfo{author}{\bibfnamefont{S.-C.} \bibnamefont{Zhang}},
  \bibinfo{journal}{Physical review letters} \textbf{\bibinfo{volume}{112}},
  \bibinfo{pages}{226801} (\bibinfo{year}{2014}).

\bibitem[{\citenamefont{Ghaemi et~al.}(2010)\citenamefont{Ghaemi, Mong, and
  Moore}}]{ghaemi2010plane}
\bibinfo{author}{\bibfnamefont{P.}~\bibnamefont{Ghaemi}},
  \bibinfo{author}{\bibfnamefont{R.~S.} \bibnamefont{Mong}}, \bibnamefont{and}
  \bibinfo{author}{\bibfnamefont{J.~E.} \bibnamefont{Moore}},
  \bibinfo{journal}{Physical review letters} \textbf{\bibinfo{volume}{105}},
  \bibinfo{pages}{166603} (\bibinfo{year}{2010}).

\bibitem[{\citenamefont{Kim et~al.}(2015)\citenamefont{Kim, Baik, Ryu, Sohn,
  Park, Park, Denlinger, Yi, Choi, and Kim}}]{kim2015observation}
\bibinfo{author}{\bibfnamefont{J.}~\bibnamefont{Kim}},
  \bibinfo{author}{\bibfnamefont{S.~S.} \bibnamefont{Baik}},
  \bibinfo{author}{\bibfnamefont{S.~H.} \bibnamefont{Ryu}},
  \bibinfo{author}{\bibfnamefont{Y.}~\bibnamefont{Sohn}},
  \bibinfo{author}{\bibfnamefont{S.}~\bibnamefont{Park}},
  \bibinfo{author}{\bibfnamefont{B.-G.} \bibnamefont{Park}},
  \bibinfo{author}{\bibfnamefont{J.}~\bibnamefont{Denlinger}},
  \bibinfo{author}{\bibfnamefont{Y.}~\bibnamefont{Yi}},
  \bibinfo{author}{\bibfnamefont{H.~J.} \bibnamefont{Choi}}, \bibnamefont{and}
  \bibinfo{author}{\bibfnamefont{K.~S.} \bibnamefont{Kim}},
  \bibinfo{journal}{Science} \textbf{\bibinfo{volume}{349}},
  \bibinfo{pages}{723} (\bibinfo{year}{2015}).

\bibitem[{\citenamefont{Baik et~al.}(2015)\citenamefont{Baik, Kim, Yi, and
  Choi}}]{baik2015emergence}
\bibinfo{author}{\bibfnamefont{S.~S.} \bibnamefont{Baik}},
  \bibinfo{author}{\bibfnamefont{K.~S.} \bibnamefont{Kim}},
  \bibinfo{author}{\bibfnamefont{Y.}~\bibnamefont{Yi}}, \bibnamefont{and}
  \bibinfo{author}{\bibfnamefont{H.~J.} \bibnamefont{Choi}},
  \bibinfo{journal}{Nano letters} \textbf{\bibinfo{volume}{15}},
  \bibinfo{pages}{7788} (\bibinfo{year}{2015}).

\bibitem[{\citenamefont{Ling et~al.}(2015)\citenamefont{Ling, Wang, Huang, Xia,
  and Dresselhaus}}]{ling2015renaissance}
\bibinfo{author}{\bibfnamefont{X.}~\bibnamefont{Ling}},
  \bibinfo{author}{\bibfnamefont{H.}~\bibnamefont{Wang}},
  \bibinfo{author}{\bibfnamefont{S.}~\bibnamefont{Huang}},
  \bibinfo{author}{\bibfnamefont{F.}~\bibnamefont{Xia}}, \bibnamefont{and}
  \bibinfo{author}{\bibfnamefont{M.~S.} \bibnamefont{Dresselhaus}},
  \bibinfo{journal}{Proceedings of the National Academy of Sciences}
  \textbf{\bibinfo{volume}{112}}, \bibinfo{pages}{4523} (\bibinfo{year}{2015}).

\bibitem[{\citenamefont{Abrikosov}(2017)}]{abrikosov2017fundamentals}
\bibinfo{author}{\bibfnamefont{A.}~\bibnamefont{Abrikosov}},
  \emph{\bibinfo{title}{Fundamentals of the Theory of Metals}}
  (\bibinfo{publisher}{Courier Dover Publications}, \bibinfo{year}{2017}).

\bibitem[{\citenamefont{Minarelli et~al.}(2019)\citenamefont{Minarelli,
  P{\"o}yh{\"o}nen, Van~Dalum, Ojanen, and Fritz}}]{minarelli2019engineering}
\bibinfo{author}{\bibfnamefont{E.~L.} \bibnamefont{Minarelli}},
  \bibinfo{author}{\bibfnamefont{K.}~\bibnamefont{P{\"o}yh{\"o}nen}},
  \bibinfo{author}{\bibfnamefont{G.~A.} \bibnamefont{Van~Dalum}},
  \bibinfo{author}{\bibfnamefont{T.}~\bibnamefont{Ojanen}}, \bibnamefont{and}
  \bibinfo{author}{\bibfnamefont{L.}~\bibnamefont{Fritz}},
  \bibinfo{journal}{Physical Review B} \textbf{\bibinfo{volume}{99}},
  \bibinfo{pages}{165413} (\bibinfo{year}{2019}).

\bibitem[{\citenamefont{Sticlet et~al.}(2012)\citenamefont{Sticlet,
  Pi{\'e}chon, Fuchs, Kalugin, and Simon}}]{sticlet2012geometrical}
\bibinfo{author}{\bibfnamefont{D.}~\bibnamefont{Sticlet}},
  \bibinfo{author}{\bibfnamefont{F.}~\bibnamefont{Pi{\'e}chon}},
  \bibinfo{author}{\bibfnamefont{J.-N.} \bibnamefont{Fuchs}},
  \bibinfo{author}{\bibfnamefont{P.}~\bibnamefont{Kalugin}}, \bibnamefont{and}
  \bibinfo{author}{\bibfnamefont{P.}~\bibnamefont{Simon}},
  \bibinfo{journal}{Physical Review B} \textbf{\bibinfo{volume}{85}},
  \bibinfo{pages}{165456} (\bibinfo{year}{2012}).

\bibitem[{\citenamefont{Kok and Lovett}(2010)}]{kok2010introduction}
\bibinfo{author}{\bibfnamefont{P.}~\bibnamefont{Kok}} \bibnamefont{and}
  \bibinfo{author}{\bibfnamefont{B.~W.} \bibnamefont{Lovett}},
  \emph{\bibinfo{title}{Introduction to optical quantum information
  processing}} (\bibinfo{publisher}{Cambridge university press},
  \bibinfo{year}{2010}).

\bibitem[{\citenamefont{Bukov et~al.}(2015)\citenamefont{Bukov, D'Alessio, and
  Polkovnikov}}]{bukov2015universal}
\bibinfo{author}{\bibfnamefont{M.}~\bibnamefont{Bukov}},
  \bibinfo{author}{\bibfnamefont{L.}~\bibnamefont{D'Alessio}},
  \bibnamefont{and}
  \bibinfo{author}{\bibfnamefont{A.}~\bibnamefont{Polkovnikov}},
  \bibinfo{journal}{Advances in Physics} \textbf{\bibinfo{volume}{64}},
  \bibinfo{pages}{139} (\bibinfo{year}{2015}).

\bibitem[{\citenamefont{Kohler et~al.}(2005)\citenamefont{Kohler, Lehmann, and
  H{\"a}nggi}}]{kohler2005driven}
\bibinfo{author}{\bibfnamefont{S.}~\bibnamefont{Kohler}},
  \bibinfo{author}{\bibfnamefont{J.}~\bibnamefont{Lehmann}}, \bibnamefont{and}
  \bibinfo{author}{\bibfnamefont{P.}~\bibnamefont{H{\"a}nggi}},
  \bibinfo{journal}{Physics Reports} \textbf{\bibinfo{volume}{406}},
  \bibinfo{pages}{379} (\bibinfo{year}{2005}).

\bibitem[{\citenamefont{Tannor}(2007)}]{tannor2007introduction}
\bibinfo{author}{\bibfnamefont{D.~J.} \bibnamefont{Tannor}},
  \emph{\bibinfo{title}{Introduction to quantum mechanics: a time-dependent
  perspective}} (\bibinfo{publisher}{University Science Books},
  \bibinfo{year}{2007}).

\bibitem[{\citenamefont{De~Giovannini and H{\"u}bener}(2019)}]{de2019floquet}
\bibinfo{author}{\bibfnamefont{U.}~\bibnamefont{De~Giovannini}}
  \bibnamefont{and}
  \bibinfo{author}{\bibfnamefont{H.}~\bibnamefont{H{\"u}bener}},
  \bibinfo{journal}{Journal of Physics: Materials}
  \textbf{\bibinfo{volume}{3}}, \bibinfo{pages}{012001} (\bibinfo{year}{2019}).

\bibitem[{\citenamefont{Eckardt and Anisimovas}(2015)}]{eckardt2015high}
\bibinfo{author}{\bibfnamefont{A.}~\bibnamefont{Eckardt}} \bibnamefont{and}
  \bibinfo{author}{\bibfnamefont{E.}~\bibnamefont{Anisimovas}},
  \bibinfo{journal}{New journal of physics} \textbf{\bibinfo{volume}{17}},
  \bibinfo{pages}{093039} (\bibinfo{year}{2015}).

\bibitem[{\citenamefont{L{\'o}pez et~al.}(2015)\citenamefont{L{\'o}pez, Scholz,
  Santos, and Schliemann}}]{lopez2015photoinduced}
\bibinfo{author}{\bibfnamefont{A.}~\bibnamefont{L{\'o}pez}},
  \bibinfo{author}{\bibfnamefont{A.}~\bibnamefont{Scholz}},
  \bibinfo{author}{\bibfnamefont{B.}~\bibnamefont{Santos}}, \bibnamefont{and}
  \bibinfo{author}{\bibfnamefont{J.}~\bibnamefont{Schliemann}},
  \bibinfo{journal}{Physical Review B} \textbf{\bibinfo{volume}{91}},
  \bibinfo{pages}{125105} (\bibinfo{year}{2015}).

\bibitem[{\citenamefont{Sengupta and Rakheja}(2017)}]{sengupta2017anisotropy}
\bibinfo{author}{\bibfnamefont{P.}~\bibnamefont{Sengupta}} \bibnamefont{and}
  \bibinfo{author}{\bibfnamefont{S.}~\bibnamefont{Rakheja}},
  \bibinfo{journal}{Applied Physics Letters} \textbf{\bibinfo{volume}{111}},
  \bibinfo{pages}{161902} (\bibinfo{year}{2017}).

\bibitem[{\citenamefont{Haldane}(1988)}]{haldane1988model}
\bibinfo{author}{\bibfnamefont{F.~D.~M.} \bibnamefont{Haldane}},
  \bibinfo{journal}{Physical review letters} \textbf{\bibinfo{volume}{61}},
  \bibinfo{pages}{2015} (\bibinfo{year}{1988}).

\bibitem[{\citenamefont{D’Alessio and Rigol}(2015)}]{d2015dynamical}
\bibinfo{author}{\bibfnamefont{L.}~\bibnamefont{D’Alessio}} \bibnamefont{and}
  \bibinfo{author}{\bibfnamefont{M.}~\bibnamefont{Rigol}},
  \bibinfo{journal}{Nature communications} \textbf{\bibinfo{volume}{6}},
  \bibinfo{pages}{1} (\bibinfo{year}{2015}).

\bibitem[{\citenamefont{Fruchart and
  Carpentier}(2013)}]{fruchart2013introduction}
\bibinfo{author}{\bibfnamefont{M.}~\bibnamefont{Fruchart}} \bibnamefont{and}
  \bibinfo{author}{\bibfnamefont{D.}~\bibnamefont{Carpentier}},
  \bibinfo{journal}{Comptes Rendus Physique} \textbf{\bibinfo{volume}{14}},
  \bibinfo{pages}{779} (\bibinfo{year}{2013}).

\bibitem[{\citenamefont{Landau et~al.}(2013)\citenamefont{Landau, Bell,
  Kearsley, Pitaevskii, Lifshitz, and Sykes}}]{landau2013electrodynamics}
\bibinfo{author}{\bibfnamefont{L.~D.} \bibnamefont{Landau}},
  \bibinfo{author}{\bibfnamefont{J.}~\bibnamefont{Bell}},
  \bibinfo{author}{\bibfnamefont{M.}~\bibnamefont{Kearsley}},
  \bibinfo{author}{\bibfnamefont{L.}~\bibnamefont{Pitaevskii}},
  \bibinfo{author}{\bibfnamefont{E.}~\bibnamefont{Lifshitz}}, \bibnamefont{and}
  \bibinfo{author}{\bibfnamefont{J.}~\bibnamefont{Sykes}},
  \emph{\bibinfo{title}{Electrodynamics of continuous media}},
  vol.~\bibinfo{volume}{8} (\bibinfo{publisher}{elsevier},
  \bibinfo{year}{2013}).

\bibitem[{\citenamefont{Yokoyama and Murakami}(2011)}]{yokoyama2011transverse}
\bibinfo{author}{\bibfnamefont{T.}~\bibnamefont{Yokoyama}} \bibnamefont{and}
  \bibinfo{author}{\bibfnamefont{S.}~\bibnamefont{Murakami}},
  \bibinfo{journal}{Physical Review B} \textbf{\bibinfo{volume}{83}},
  \bibinfo{pages}{161407} (\bibinfo{year}{2011}).

\bibitem[{\citenamefont{Bergman and Oganesyan}(2010)}]{bergman2010theory}
\bibinfo{author}{\bibfnamefont{D.~L.} \bibnamefont{Bergman}} \bibnamefont{and}
  \bibinfo{author}{\bibfnamefont{V.}~\bibnamefont{Oganesyan}},
  \bibinfo{journal}{Physical review letters} \textbf{\bibinfo{volume}{104}},
  \bibinfo{pages}{066601} (\bibinfo{year}{2010}).

\bibitem[{\citenamefont{Xiao et~al.}(2006)\citenamefont{Xiao, Yao, Fang, and
  Niu}}]{xiao2006berry}
\bibinfo{author}{\bibfnamefont{D.}~\bibnamefont{Xiao}},
  \bibinfo{author}{\bibfnamefont{Y.}~\bibnamefont{Yao}},
  \bibinfo{author}{\bibfnamefont{Z.}~\bibnamefont{Fang}}, \bibnamefont{and}
  \bibinfo{author}{\bibfnamefont{Q.}~\bibnamefont{Niu}},
  \bibinfo{journal}{Physical review letters} \textbf{\bibinfo{volume}{97}},
  \bibinfo{pages}{026603} (\bibinfo{year}{2006}).

\bibitem[{\citenamefont{Rakheja and Sengupta}(2016)}]{rakheja2016tuning}
\bibinfo{author}{\bibfnamefont{S.}~\bibnamefont{Rakheja}} \bibnamefont{and}
  \bibinfo{author}{\bibfnamefont{P.}~\bibnamefont{Sengupta}},
  \bibinfo{journal}{Journal of Physics D: Applied Physics}
  \textbf{\bibinfo{volume}{49}}, \bibinfo{pages}{115106}
  (\bibinfo{year}{2016}).

\bibitem[{\citenamefont{Li et~al.}(2017{\natexlab{a}})\citenamefont{Li, Xu, Ba,
  Xuan, Chen, Sun, Zhang, and Zhang}}]{li2017tunable}
\bibinfo{author}{\bibfnamefont{D.}~\bibnamefont{Li}},
  \bibinfo{author}{\bibfnamefont{J.-R.} \bibnamefont{Xu}},
  \bibinfo{author}{\bibfnamefont{K.}~\bibnamefont{Ba}},
  \bibinfo{author}{\bibfnamefont{N.}~\bibnamefont{Xuan}},
  \bibinfo{author}{\bibfnamefont{M.}~\bibnamefont{Chen}},
  \bibinfo{author}{\bibfnamefont{Z.}~\bibnamefont{Sun}},
  \bibinfo{author}{\bibfnamefont{Y.-Z.} \bibnamefont{Zhang}}, \bibnamefont{and}
  \bibinfo{author}{\bibfnamefont{Z.}~\bibnamefont{Zhang}}, \bibinfo{journal}{2D
  Materials} \textbf{\bibinfo{volume}{4}}, \bibinfo{pages}{031009}
  (\bibinfo{year}{2017}{\natexlab{a}}).

\bibitem[{\citenamefont{Liu et~al.}(2017)\citenamefont{Liu, Qiu, Carvalho, Bao,
  Xu, Tan, Liu, Castro~Neto, Loh, and Lu}}]{liu2017gate}
\bibinfo{author}{\bibfnamefont{Y.}~\bibnamefont{Liu}},
  \bibinfo{author}{\bibfnamefont{Z.}~\bibnamefont{Qiu}},
  \bibinfo{author}{\bibfnamefont{A.}~\bibnamefont{Carvalho}},
  \bibinfo{author}{\bibfnamefont{Y.}~\bibnamefont{Bao}},
  \bibinfo{author}{\bibfnamefont{H.}~\bibnamefont{Xu}},
  \bibinfo{author}{\bibfnamefont{S.~J.} \bibnamefont{Tan}},
  \bibinfo{author}{\bibfnamefont{W.}~\bibnamefont{Liu}},
  \bibinfo{author}{\bibfnamefont{A.}~\bibnamefont{Castro~Neto}},
  \bibinfo{author}{\bibfnamefont{K.~P.} \bibnamefont{Loh}}, \bibnamefont{and}
  \bibinfo{author}{\bibfnamefont{J.}~\bibnamefont{Lu}}, \bibinfo{journal}{Nano
  letters} \textbf{\bibinfo{volume}{17}}, \bibinfo{pages}{1970}
  (\bibinfo{year}{2017}).

\bibitem[{\citenamefont{Li et~al.}(2017{\natexlab{b}})\citenamefont{Li, Xu,
  Ding, Wang, Shen, Lu, Zhu, and Behnia}}]{li2017anomalous}
\bibinfo{author}{\bibfnamefont{X.}~\bibnamefont{Li}},
  \bibinfo{author}{\bibfnamefont{L.}~\bibnamefont{Xu}},
  \bibinfo{author}{\bibfnamefont{L.}~\bibnamefont{Ding}},
  \bibinfo{author}{\bibfnamefont{J.}~\bibnamefont{Wang}},
  \bibinfo{author}{\bibfnamefont{M.}~\bibnamefont{Shen}},
  \bibinfo{author}{\bibfnamefont{X.}~\bibnamefont{Lu}},
  \bibinfo{author}{\bibfnamefont{Z.}~\bibnamefont{Zhu}}, \bibnamefont{and}
  \bibinfo{author}{\bibfnamefont{K.}~\bibnamefont{Behnia}},
  \bibinfo{journal}{Physical Review Letters} \textbf{\bibinfo{volume}{119}},
  \bibinfo{pages}{056601} (\bibinfo{year}{2017}{\natexlab{b}}).

\bibitem[{\citenamefont{Flores et~al.}(2015)\citenamefont{Flores, Ares,
  Castellanos-Gomez, Barawi, Ferrer, and
  S{\'a}nchez}}]{flores2015thermoelectric}
\bibinfo{author}{\bibfnamefont{E.}~\bibnamefont{Flores}},
  \bibinfo{author}{\bibfnamefont{J.~R.} \bibnamefont{Ares}},
  \bibinfo{author}{\bibfnamefont{A.}~\bibnamefont{Castellanos-Gomez}},
  \bibinfo{author}{\bibfnamefont{M.}~\bibnamefont{Barawi}},
  \bibinfo{author}{\bibfnamefont{I.~J.} \bibnamefont{Ferrer}},
  \bibnamefont{and}
  \bibinfo{author}{\bibfnamefont{C.}~\bibnamefont{S{\'a}nchez}},
  \bibinfo{journal}{Applied Physics Letters} \textbf{\bibinfo{volume}{106}},
  \bibinfo{pages}{022102} (\bibinfo{year}{2015}).

\bibitem[{\citenamefont{Saito et~al.}(2016)\citenamefont{Saito, Iizuka,
  Koretsune, Arita, Shimizu, and Iwasa}}]{saito2016gate}
\bibinfo{author}{\bibfnamefont{Y.}~\bibnamefont{Saito}},
  \bibinfo{author}{\bibfnamefont{T.}~\bibnamefont{Iizuka}},
  \bibinfo{author}{\bibfnamefont{T.}~\bibnamefont{Koretsune}},
  \bibinfo{author}{\bibfnamefont{R.}~\bibnamefont{Arita}},
  \bibinfo{author}{\bibfnamefont{S.}~\bibnamefont{Shimizu}}, \bibnamefont{and}
  \bibinfo{author}{\bibfnamefont{Y.}~\bibnamefont{Iwasa}},
  \bibinfo{journal}{Nano letters} \textbf{\bibinfo{volume}{16}},
  \bibinfo{pages}{4819} (\bibinfo{year}{2016}).

\end{thebibliography}
%\bibliographystyle{apsrev}

\end{document}